\documentclass[prl, twocolumn,amsmath,floatfix]{revtex4}
\usepackage{graphicx}
\usepackage{amsfonts}

\begin{document}

\title{Coarse-Grained Model of Entropic Allostery}
\author{Rhoda J. Hawkins}
\email{rhoda.hawkins@physics.org}
\author{Thomas C.B. McLeish}
\affiliation{IRC in Polymer Science and Technology, Department of Physics and Astronomy, University of Leeds, Leeds LS2
  9JT, United Kingdom}
\date{\today}
\begin{abstract}Many signalling functions in molecular biology require proteins to bind to substrates such as DNA in response to environmental
  signals such as the simultaneous binding to a small molecule. Examples are repressor
  proteins which may transmit information via a conformational change in response to the ligand
  binding. An alternative entropic mechanism of ``allostery'' suggests
  that the inducer ligand changes the intramolecular
  vibrational entropy, not just the mean static structure. We present a
  quantitative, coarse-grained model
  of entropic allostery, which suggests design rules for internal
  cohesive potentials in proteins employing this effect. It also addresses the issue of how the signal
  information to bind or unbind is transmitted through the
  protein. The model may be applicable
  to a wide range of repressors and also to signalling in
  transmembrane proteins.
\end{abstract}
\maketitle
\textit{Introduction.} 
It is becoming increasingly clear that dynamics, as well as static structure, are important in
molecular biology. For example simulations of dynamical
transitions in proteins \cite{Tour+03} suggest that collective global modes are correlated with protein
function. This letter focuses on repressor
proteins which bind to DNA to
``turn off'' genes when the cell does not require their
expression. The binding is ``allosteric'': it is activated
depending on the presence of inducer ligands, small molecules which
themselves bind to the protein. The ``holorepressor''
(``aporepressor'') is the protein with (without) a bound
ligand. Frequently the ligand binding site on the protein is distant from that
of the DNA. For this reason ligand binding has often been assumed to cause a 
conformational change within the
repressor protein, decreasing its affinity for DNA in one state
compared to the other state. However, this is evidence that dynamically induced entropic changes may
contribute to
allostery~\cite{BellLewis01,Jusuf+03,CoopDry84}.

A classic example of a repressor system is the E-coli lac
repressor \cite{Brui02b,Alberts,Lewis96,BellLewis00,Lehn00}. In
  this case the aporepressor binds to DNA, suppressing the
genes for the 
metabolism of lactose.
A second example is the E-coli trp
repressor~\cite{Otw+88,LawCar93,Zhao+93} that, on binding, prevents
the expression of the gene for
tryptophan synthesis. But in contrast to the lac, trp-type holorepressor proteins bind to DNA and the
aporepressors do not.
There are many such repressor systems but the lac and the trp will act
as representative cases for this Letter. Our challenge is to explore whether
the Brownian fluctuations in protein structure may carry information
between the two binding sites, thereby producing cooperative lac-type
or trp-type behaviour. This mechanism of cooperativity is one of the key questions in
understanding protein function \cite{Matt96}.

\begin{figure}[bh]
  \includegraphics[width=7cm]{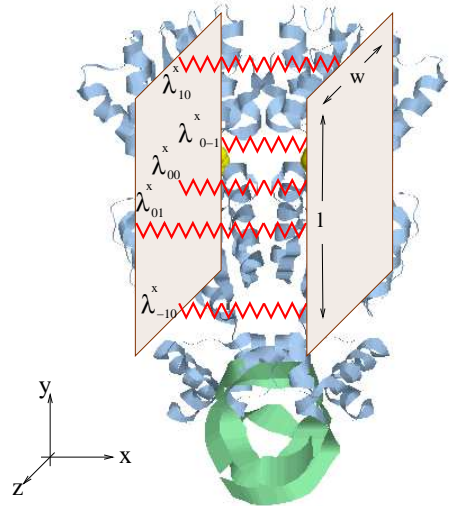}
 \caption{\label{model}Plates and springs model for the interaction of
   the two domains of a repressor dimer. The xray structure (PDB 1EFA \cite{Lewis96}) of the lac
   repressor dimer (DNA shown in green) is shown behind the model.}
\end{figure}
\textit{A coarse-grained model.}
In a coarse-grained representation we model a repressor protein dimer as two
rigid plates of length $l$ and width $w$, representing the two protein
monomers motivated by the common dimer motif (see
Figure~\ref{model}). We parameterise the relative motion of the plates by three relative translation displacements ($x$, $y$, $z$) and three rotation angles ($\theta_x$, $\theta_y$, $\theta_z$). The stabilising contact interaction between the two dimers is characterised by
5 quadratic (3D) potential wells. Figure~\ref{model} shows 5 of the
effective springs that arise, $\lambda^x_i$ which
are perpendicular to the plates. We find that a minimal model requires
just 4 other springs in the plane
of the plates (summations of diagonal springs between the plates), which we label as $\lambda^y_{00}$, $\lambda^z_{0-1}$, $\lambda^z_{00}$, $\lambda^z_{01}$, representing the resolved $y$ and $z$ relative displacements. These
local interactions represent a minimal set of ``sticky patches'' which could arise
from hydrophobic, side chain, or electrostatic
forces.  The corresponding spring
constants $\lambda_{i}$ could be calculated
in principle from the details of these
interactions. We allow only local modifications of the contact interactions on binding. 
The springs local to the ligand binding site will be
affected by the binding of an inducer or corepressor. (Lac actually
binds two inducer ligands, which we simplify here with a single bound
state.) Similarly springs local to the DNA binding site will be
affected by binding to DNA. The other springs act as anchoring potentials.

We consider vibrations of the plates in 3 different planes: (i) in the $x,y$ plane
(translational vibrations along the
$x$-axis and rotations about the $z$-axis); (ii) in
the $x,z$ plane (translation along the $x$-axis and rotation about the $y$-axis); and (iii) in the $y,z$ plane
(translation along the $z$-axis and rotation about the $x$-axis).

The Tirion potential~\cite{Tirion96} replaces the full MD potential
with a simple pairwise quadratic potential of universal strength and this is found to be
sufficient to describe the low frequency modes involving coherent
motion of large groups of atoms. In our model we similarly look only at these
low frequency modes and describe them by harmonic potentials between
the protein domains, but allow the potentials between the protein monomers to acquire locally specific values.

As an example of detailed calculations we take the $2\times 2$ system
of motions in the $y,z$ plane. We write $z_{\scriptscriptstyle 01}=z_{\scriptscriptstyle 00}-\frac{l\theta_x}{2}$ and
$z_{\scriptscriptstyle 0-1}=z_{\scriptscriptstyle 00}+\frac{l\theta_x}{2}$ (where
$\theta_x$ is the angle of rotation about the $x$-axis) to obtain the Hamiltonian
in terms of the mutual translational
vibration coordinate $z_{\scriptscriptstyle 00}$ and mutual rotational vibration
coordinate $\theta_x$.
\begin{equation}
H=\frac{1}{2}{\bf
  pM^{-1}p}+\frac{1}{2}{\bf xKx}
\end{equation}
where the interaction matrix for the $y,z$ plane
\begin{eqnarray*}
\mathbf{K}&=&\left(\begin{array}{cc}-(\lambda^z_{01}+\lambda^z_{0-1}+\lambda^z_{00}) &\frac{1}{2}(\lambda^z_{01}-\lambda^z_{0-1})\\\frac{1}{2}(\lambda^z_{01}-\lambda^z_{0-1}) &-\frac{1}{4}(\lambda^z_{01}+\lambda^z_{0-1})
  \end{array}\right),
\end{eqnarray*}
the inertial matrix
\begin{eqnarray*}
\mathbf{M}&=&\left(\begin{array}{cc}m &0\\ 0&\frac{I}{l^2}
  \end{array}\right),
\:\:\:\text{and}\:\:\:\:\:\mathbf{x}=\left(\begin{array}{c} z_{\scriptscriptstyle 00} \\l\theta_x \end{array}\right)
\end{eqnarray*}
where $m$ is the reduced mass and $I$ is the reduced moment of inertia
of the dimer pair.
This leads to the partition function in the relevant classical limit
\begin{equation}\label{Z}
Z=\int... \int e^{\frac{-H(x_{\scriptscriptstyle 0_i},\theta_i)}{kT}}\mathrm{d}x_{\scriptscriptstyle 0_i}\,\mathrm{d}\theta_i=\frac{2\pi
  kT}{(|\mathbf{M^{-1}}||\mathbf{K}|)^{1/2}}.
\end{equation}
Finally from (\ref{Z}) the entropy of the protein dimer for a single
plane is
\begin{equation}
S=Nk(\ln2\pi
kT\sqrt{mI}/l+1-1/2\ln(\lambda_1\lambda_{-1}+\lambda_0(\lambda_1+\lambda_{-1})/4))\end{equation}
where $\lambda^z_{0i}$ has been abbreviated to $\lambda_i$ for convenience.
We are interested in the difference between the change in entropy {\it
  on binding at the DNA binding site of the two cases in which the
protein is, and is not also bound to the inducer.}
We call this $\Delta\Delta
S=\Delta S_{\mathrm{holo}}-\Delta S_{\mathrm{apo}}$. A result with
$\Delta\Delta S \neq 0$ would signify cooperative behaviour i.e. the
binding to DNA is affected by the binding to the inducer. We write $\Delta\Delta S$ in terms of dimensionless spring constants
$\tilde{\lambda_1}=\frac{\lambda_1}{\lambda_0}$ and
$\tilde{\lambda_{-1}}=\frac{\lambda_{-1}}{\lambda_0}$ and bound to unbound
ratios $\Lambda_1=\frac{\lambda_{1B}}{\lambda_{1}}$ and
$\Lambda_{-1}=\frac{\lambda_{-1B}}{\lambda_{-1}}$. This gives us
\begin{equation}\label{DeltaDeltaS}
\Delta\Delta
S=\frac{1}{2}Nk\ln\left(\frac{(4\Lambda_1+\frac{\Lambda_1}{\tilde{\lambda_{-1}}}+\frac{1}{\tilde{\lambda_1}})(4\Lambda_{-1}+\frac{\Lambda_{-1}}{\tilde{\lambda_1}}+\frac{1}{\tilde{\lambda_{-1}}})}{(4\Lambda_1\Lambda_{-1}+\frac{\Lambda_1}{\tilde{\lambda_{-1}}}+\frac{\Lambda_{-1}}{\tilde{\lambda_1}})(4+\frac{1}{\tilde{\lambda_{-1}}}+\frac{1}{\tilde{\lambda_1}})}\right).
\end{equation}
The other modes can be modelled in the same way to give additional contributions to $\Delta\Delta S$.
The 2-plate model generates
a $3\times 3$ form of ${\mathbf M}$ and ${\mathbf K}$ for coupled
rotations about $y$ and $z$ and translations along $x$ plus one simple
$y$-translation.

We take $N=2$ for the lac since it is a tetramer of
two dimers and similarly for the trp since it represses as a dimer of
dimers so also has two dimers.

$\Delta\Delta S>0$ gives the trp case whereby  the affinity for
the holorepressor binding to DNA is greater than the apo
repressor. $\Delta\Delta S<0$ however gives the lac case since the apo lac repressor is
the one with the higher affinity for DNA.
Applying these inequalities to equation~(\ref{DeltaDeltaS})  gives the
following rule determining which case arises.
\begin{equation}\Lambda_1\Lambda_{-1}+1\left\{ \begin{array}{c}>\\< \end{array}\right\}\Lambda_1+\Lambda_{-1} \:\:\:\:\:\:\left\{\begin{array}{c}
  \mathrm{trp} \\
  \mathrm{lac}\end{array}\right\}.
\end{equation}
The trp case occurs when both spring constants increase
($\Lambda_1>1$, $\Lambda_{-1}>1$) or decrease ($\Lambda_1<1$,
$\Lambda_{-1}<1$) on ligand binding. The lac case occurs when one spring
constant increases and the other decreases ($\Lambda_1>1$ and
$\Lambda_{-1}<1$ or $\Lambda_1<1$ and $\Lambda_{-1}>1$).
\begin{figure}[!hbt]
   \includegraphics[width=8cm]{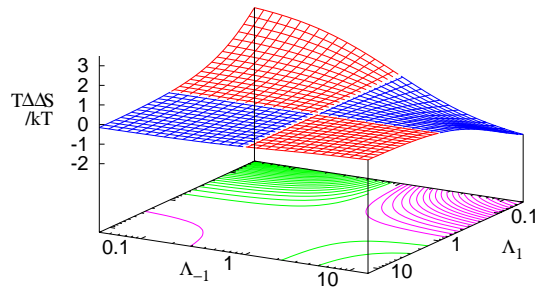}
\caption{\label{3D}Graph showing $\Delta\Delta S$ against $\Lambda_1$
  and $\Lambda_{-1}$. The contour plot on the base shows regions where lac-type (magenta) and trp-type (green) behaviour is optimised.}
\end{figure}
Figure~\ref{3D} plots the function $\Delta\Delta
S(\Lambda_1,\Lambda_{-1})$ (equation~\ref{DeltaDeltaS}). Biologically
relevant values for the original spring constants were chosen using
protein $B$-factor data (related to the rms positions of the atoms $B=8\pi^2<\bar{u^2}>$ \cite{Rhodes}) and steered simulations (see later) giving the case for a potential
which is stronger at the inducer binding site ($\tilde{\lambda_1}>\tilde{\lambda_{-1}}$).
It can be seen that the negative $\Delta\Delta S$ lac-type effect is
maximised when the ligand binding at the inducer binding site
decreases the spring constant $\tilde{\lambda_1}$ but $\tilde{\lambda_{-1}}$ increases on
DNA binding. The trp-type effect (positive $\Delta\Delta S$) is maximised
when both spring constants decrease as much as possible on
binding. This reduction will however be limited by the physical requirement for overall
stability of the complex. Requiring that the RMS displacement of the monomers be less than
the average separation of the atoms leads to the estimation that
$\lambda_i>0.1kT$\AA$^{-2}$. 

Physically, ``entropic allostery'' allows the lac inducer binding to communicate via the large amplitudes of the internal modes
of the protein to the ``read-head'' binding regions near the DNA which as a
result move too much to be inserted into the DNA.

\textit{Estimating the spring constants.} 
We evaluate our model in the real example of the lac repressor as an illustration.
Firstly we simply calculated estimates for the spring constants
from the x-ray crystal $B$-factors of the solved structures in the
protein data bank. 
From the solved structure \cite{Lewis96,Fried+95,BellLewis00} it is clear that $\lambda_1$ decreases
on inducer binding since the headpiece domains are disrupted on lac
binding. On DNA binding specific
interactions are formed between the repressor headpiece
domains and the base pairs in the DNA major groove which would increase the
local binding potential and so increase $\lambda_{-1}$. These observations therefore
support our finding that the lac case has $\Lambda_1<1$ and $\Lambda_{-1}>1$.

We converted the protein data for the $B$-factors \cite{Lewis96,Fried+95,BellLewis00}
into RMS vibration values and estimated the
spring constants $\lambda_i$ and $\Lambda_i$ (in this case averaged
over the vibrations in the different planes) from the expression $\lambda \sim 1/<\bar{u}^2>$
giving $\tilde{\lambda_1} \approx 1.2$, $\tilde{\lambda_{-1}}\approx 0.1$ (estimated from
\cite{Slij+97}), $\Lambda_1 \approx 0.07$ and $\Lambda_{-1} \approx 6.7$.
We now calculate an estimate of $\Delta\Delta S$ using equation
(\ref{DeltaDeltaS}). Including a factor of three due to the three planes of vibration, we obtain for our plate-dimer model a value of
$T\Delta\Delta S \sim -1.4kT$. Since the experimental values for the change in binding energy between holo and
aporepressor binding to DNA are $\Delta\Delta G \sim 6kT$
\cite{BarkBour80,HortLewLu97,Leva+96} this indicates that the
entropic contribution is likely to be significant since the crystal
dynamics can only be a lower bound for the amplitudes of vibration.

To improve upon this rough estimate we
calculated the interaction energy between the lac monomers in a fully
atomistic computation using the software ``cns'' \cite{cns} and steering the relative positions of the two monomers.
We used the xray crystal structure coordinates PDB ID 1LBI \cite{Lewis96} for the
aporepressor and 1TLF \cite{Fried+95} for the holorepressor.
\begin{figure}[!hbt]
   \includegraphics[width=6cm]{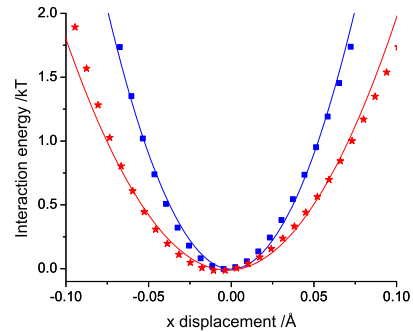}
\caption{\label{wells} Graph showing an example of a change in spring
  constant with and without the inducer for the lac repressor. The blue squares are cns data for 1LBI the aporepressor and the red stars are cns data for 1TLF the holorepressor. The lines are quadratic fits.}
\end{figure}
By relative translation along the 3 axes, and rotation about axes at
the extremities of the protein dimer, and 
recalculating the total interaction energy at each increment, we were able to build up curves for the potential energy
wells (see Figure \ref{wells}). By curve fitting the bottom of these
wells to a quadratic (to
fit with our harmonic approximation) we were able to extract the
curvature and therefore the spring constants for each global mode
of both the holo and aporepressors. We then used these  to calculate $\Delta S_{\mathrm{inducer bind}}$
which is an estimate for $-\Delta\Delta S$ in the case of large $\Lambda_{-1}$. For the motion in the $x,y$ and $x,z$ planes (translation along $x$ and rotation about $y$ and $z$) this gave $T\Delta\Delta S
\sim -1.66kT$, for the $y,z$ plane $-0.42kT$ and for the relative translation along $y$ (which does not have rotation modes coupled to it)
$-0.28kT$. Interestingly the softest mode, contributing most to the allostery,
is the one which shifts the DNA read heads (which point in the plane
perpendicular to the core) away from the DNA
perpendicular to it (the $x,z$ plane). Therefore in total we have an estimate for $\Delta\Delta S\sim
-2.36kT$.
\begin{figure}[!hbt]
   \includegraphics[width=7cm]{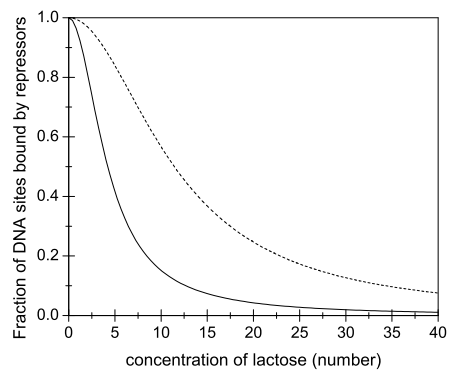}
\caption{\label{probb} Graph showing the fraction of DNA sites
  bound by repressors against the concentration of inducer
  lactose. The solid curve includes the entropic component and the
  dashed curve is without the entropic component.}
\end{figure}

To check how significant this entropy contribution is to the total
$\Delta\Delta G\sim 6kT$ we compared the probability of genes
being repressed (bound by repressor) against lactose concentration
with and without this entropic
contribution (following Yildirim \cite{Yild+03})(see Figure~\ref{probb}). For $95\%$ activation (operators not bound) 18 lactose
molecules are required with the entropic contribution but 50 would be
required if there was no entropic contribution. Note there are only of
order 100 repressors in the cell \cite{Brui02b}. This would imply that the
entropic contribution to allostery is significant in controlling the lactose level
at which the gene expression is turned on.

\textit{Conclusions.} 
We conclude that inducer binding affects the Brownian motion within
the repressor protein and this entropic effect contributes to the
allosteric mechanism in DNA binding proteins alongside static
conformational changes. We can relate the
communication of the signal across the protein to ``design rules'' for
the potentials within it.

Several extensions of this approach suggest themselves. A discrete
many-springs model naturally extends to the case of a continuous
potential between the plates. Secondly, the case of multiple,
sequential ligand binding will lead to additional structure. To make the
model even more realistic we should also include bending modes of the
protein monomers themselves. Significantly, addition of such bending modes must
increase the predicted $\Delta\Delta S$ value if the ligand binding
changes the bending rigidity. For lac any such
increase might give values of $\Delta \Delta S$ that actually dominate
the binding free energy.

We expect this model can be also applied to
transmembrane proteins that transmit signals across membranes into
cells and organelles \cite{Spie+91, Swin+03}. These systems are similar in that an inducer
ligand (e.g. adrenaline) binds to the receptor transmembrane protein which in turn
allows it to bind or unbind proteins on the interior of
the membrane (e.g. to bind to a G-protein in the control cycle for
glycogen). Within the restricted environment of the
membrane we expect entropic allostery to play an important role in the
transmission of the signal through the receptor protein.

The calculated values of the contribution to the free
energy change from the change in intramolecular vibrational entropy of
the protein easily reach the order of a few $kT$ per molecule, within
the experimentally observed range for these systems.

\begin{acknowledgments}
We thank the EPSRC for funding, and P. G. Stockley, S. E. V. Phillips,
S. W. Homans and R. F. Bruinsma for stimulating discussions.
\end{acknowledgments}


\begin{thebibliography}{24}
\expandafter\ifx\csname natexlab\endcsname\relax\def\natexlab#1{#1}\fi
\expandafter\ifx\csname bibnamefont\endcsname\relax
  \def\bibnamefont#1{#1}\fi
\expandafter\ifx\csname bibfnamefont\endcsname\relax
  \def\bibfnamefont#1{#1}\fi
\expandafter\ifx\csname citenamefont\endcsname\relax
  \def\citenamefont#1{#1}\fi
\expandafter\ifx\csname url\endcsname\relax
  \def\url#1{\texttt{#1}}\fi
\expandafter\ifx\csname urlprefix\endcsname\relax\def\urlprefix{URL }\fi
\providecommand{\bibinfo}[2]{#2}
\providecommand{\eprint}[2][]{\url{#2}}

\bibitem[{\citenamefont{Tournier and Smith}(2003)}]{Tour+03}
\bibinfo{author}{\bibfnamefont{A.~L.} \bibnamefont{Tournier}} \bibnamefont{and}
  \bibinfo{author}{\bibfnamefont{J.~C.} \bibnamefont{Smith}},
  \bibinfo{journal}{Phys. Rev. Lett.} \textbf{\bibinfo{volume}{91}},
  \bibinfo{pages}{208106} (\bibinfo{year}{2003}).

\bibitem[{\citenamefont{Bell and Lewis}(2001)}]{BellLewis01}
\bibinfo{author}{\bibfnamefont{C.~E.} \bibnamefont{Bell}} \bibnamefont{and}
  \bibinfo{author}{\bibfnamefont{M.}~\bibnamefont{Lewis}},
  \bibinfo{journal}{Current Opinion in Structural Biology}
  \textbf{\bibinfo{volume}{11}}, \bibinfo{pages}{19} (\bibinfo{year}{2001}).

\bibitem[{\citenamefont{Jusuf et~al.}(2003)\citenamefont{Jusuf, Loll, and
  Axelsen}}]{Jusuf+03}
\bibinfo{author}{\bibfnamefont{S.}~\bibnamefont{Jusuf}},
  \bibinfo{author}{\bibfnamefont{P.~J.} \bibnamefont{Loll}}, \bibnamefont{and}
  \bibinfo{author}{\bibfnamefont{P.~H.} \bibnamefont{Axelsen}},
  \bibinfo{journal}{Journal of the American Chemical Society}
  \textbf{\bibinfo{volume}{125}}, \bibinfo{pages}{3988} (\bibinfo{year}{2003}).

\bibitem[{\citenamefont{Cooper and Dryden}(1984)}]{CoopDry84}
\bibinfo{author}{\bibfnamefont{A.}~\bibnamefont{Cooper}} \bibnamefont{and}
  \bibinfo{author}{\bibfnamefont{D.~T.~F.} \bibnamefont{Dryden}},
  \bibinfo{journal}{European Biophysics Journal} \textbf{\bibinfo{volume}{11}}
  (\bibinfo{year}{1984}).

\bibitem[{\citenamefont{Lewis et~al.}(1996)\citenamefont{Lewis, Chang, Horton,
  Kercher, Pace, Schumacher, Brennan, and Lu}}]{Lewis96}
\bibinfo{author}{\bibfnamefont{M.}~\bibnamefont{Lewis}},
  \bibinfo{author}{\bibfnamefont{G.}~\bibnamefont{Chang}},
  \bibinfo{author}{\bibfnamefont{N.~C.} \bibnamefont{Horton}},
  \bibinfo{author}{\bibfnamefont{M.~A.} \bibnamefont{Kercher}},
  \bibinfo{author}{\bibfnamefont{H.~C.} \bibnamefont{Pace}},
  \bibinfo{author}{\bibfnamefont{M.~A.} \bibnamefont{Schumacher}},
  \bibinfo{author}{\bibfnamefont{R.~G.} \bibnamefont{Brennan}},
  \bibnamefont{and} \bibinfo{author}{\bibfnamefont{P.}~\bibnamefont{Lu}},
  \bibinfo{journal}{Science} \textbf{\bibinfo{volume}{271}},
  \bibinfo{pages}{1247} (\bibinfo{year}{1996}), \bibinfo{note}{-PDB ID: 1LBI}.


\bibitem[{\citenamefont{Bruinsma}(2002)}]{Brui02b}
\bibinfo{author}{\bibfnamefont{R.~F.} \bibnamefont{Bruinsma}},
  \bibinfo{journal}{Physica A} \textbf{\bibinfo{volume}{313}},
  \bibinfo{pages}{211} (\bibinfo{year}{2002}).

\bibitem[{\citenamefont{Alberts}(2002)}]{Alberts}
\bibinfo{author}{\bibfnamefont{B.}~\bibnamefont{Alberts}},
  \emph{\bibinfo{title}{Molecular Biology of the cell}}
  (\bibinfo{publisher}{Garland Science}, \bibinfo{year}{2002}),
  \bibinfo{edition}{4th} ed.

\bibitem[{\citenamefont{Bell and Lewis}(2000)}]{BellLewis00}
\bibinfo{author}{\bibfnamefont{C.~E.} \bibnamefont{Bell}} \bibnamefont{and}
  \bibinfo{author}{\bibfnamefont{M.}~\bibnamefont{Lewis}},
  \bibinfo{journal}{Nature Structural Biology} \textbf{\bibinfo{volume}{7}},
  \bibinfo{pages}{209} (\bibinfo{year}{2000}), \bibinfo{note}{-PDB ID: 1EFA}.

\bibitem[{\citenamefont{Lehninger et~al.}(2000)\citenamefont{Lehninger, Nelson,
  and Cox}}]{Lehn00}
\bibinfo{author}{\bibfnamefont{A.~L.} \bibnamefont{Lehninger}},
  \bibinfo{author}{\bibfnamefont{D.~L.} \bibnamefont{Nelson}},
  \bibnamefont{and} \bibinfo{author}{\bibfnamefont{M.~M.} \bibnamefont{Cox}},
  \emph{\bibinfo{title}{\sloppy Lehninger Principles of Biochemistry}}
  (\bibinfo{publisher}{Worth}, \bibinfo{year}{2000}), \bibinfo{edition}{3rd}
  ed., \bibinfo{note}{biochemistry in 3D Tutorial website
  www.worthpublishers.com/lehninger3D}.

\bibitem[{\citenamefont{Otwinowski et~al.}(1988)\citenamefont{Otwinowski,
  Schevitz, Zhang, Lawson, Joachimak, Marmorstein, Luisi, and Sigler}}]{Otw+88}
\bibinfo{author}{\bibfnamefont{L.}~\bibnamefont{Otwinowski}},
  \bibinfo{author}{\bibfnamefont{R.~W.} \bibnamefont{Schevitz}},
  \bibinfo{author}{\bibfnamefont{R.-G.} \bibnamefont{Zhang}},
  \bibinfo{author}{\bibfnamefont{C.~L.} \bibnamefont{Lawson}},
  \bibinfo{author}{\bibfnamefont{A.}~\bibnamefont{Joachimak}},
  \bibinfo{author}{\bibfnamefont{R.~Q.} \bibnamefont{Marmorstein}},
  \bibinfo{author}{\bibfnamefont{B.~F.} \bibnamefont{Luisi}}, \bibnamefont{and}
  \bibinfo{author}{\bibfnamefont{P.}~\bibnamefont{Sigler}},
  \bibinfo{journal}{Nature} \textbf{\bibinfo{volume}{335}},
  \bibinfo{pages}{321} (\bibinfo{year}{1988}).

\bibitem[{\citenamefont{Lawson and Carey}(1993)}]{LawCar93}
\bibinfo{author}{\bibfnamefont{C.~L.} \bibnamefont{Lawson}} \bibnamefont{and}
  \bibinfo{author}{\bibfnamefont{J.}~\bibnamefont{Carey}},
  \bibinfo{journal}{Nature} \textbf{\bibinfo{volume}{366}},
  \bibinfo{pages}{178} (\bibinfo{year}{1993}), \bibinfo{note}{-PDB ID 1TRR}.

\bibitem[{\citenamefont{Zhao et~al.}(1993)\citenamefont{Zhao, Arrowsmith, Jia,
  and Jardetzky}}]{Zhao+93}
\bibinfo{author}{\bibfnamefont{D.}~\bibnamefont{Zhao}},
  \bibinfo{author}{\bibfnamefont{C.~H.} \bibnamefont{Arrowsmith}},
  \bibinfo{author}{\bibfnamefont{X.}~\bibnamefont{Jia}}, \bibnamefont{and}
  \bibinfo{author}{\bibfnamefont{O.}~\bibnamefont{Jardetzky}},
  \bibinfo{journal}{Journal of molecular biology}
  \textbf{\bibinfo{volume}{229}}, \bibinfo{pages}{735} (\bibinfo{year}{1993}),
  \bibinfo{note}{-PDB ID: 1WRS and 1WRT}.

\bibitem[{\citenamefont{Matthews}(1996)}]{Matt96}
\bibinfo{author}{\bibfnamefont{K.~S.} \bibnamefont{Matthews}},
  \bibinfo{journal}{Science} \textbf{\bibinfo{volume}{271}},
  \bibinfo{pages}{1245} (\bibinfo{year}{1996}).

\bibitem[{\citenamefont{Tirion}(1996)}]{Tirion96}
\bibinfo{author}{\bibfnamefont{M.~M.} \bibnamefont{Tirion}},
  \bibinfo{journal}{Phys. Rev. Lett.} \textbf{\bibinfo{volume}{77}},
  \bibinfo{pages}{1905} (\bibinfo{year}{1996}).

\bibitem[{\citenamefont{Friedman et~al.}(1995)\citenamefont{Friedman,
  Fischmann, and Steitz}}]{Fried+95}
\bibinfo{author}{\bibfnamefont{A.~M.} \bibnamefont{Friedman}},
  \bibinfo{author}{\bibfnamefont{T.~O.} \bibnamefont{Fischmann}},
  \bibnamefont{and} \bibinfo{author}{\bibfnamefont{T.~A.}
  \bibnamefont{Steitz}}, \bibinfo{journal}{Science}
  \textbf{\bibinfo{volume}{268}}, \bibinfo{pages}{1721} (\bibinfo{year}{1995}),
  \bibinfo{note}{-PDB ID: 1TLF}.

\bibitem[{\citenamefont{Rhodes}(1993)}]{Rhodes}
\bibinfo{author}{\bibfnamefont{G.}~\bibnamefont{Rhodes}},
  \emph{\bibinfo{title}{Crystallography made crystal clear}}
  (\bibinfo{publisher}{Academic Press Inc.}, \bibinfo{address}{San Diego,
  California}, \bibinfo{year}{1993}).

\bibitem[{\citenamefont{Slijper et~al.}(1997)\citenamefont{Slijper, Boelens,
  Davis, Konings, van~der Marel, van Boom, and Kaptein}}]{Slij+97}
\bibinfo{author}{\bibfnamefont{M.}~\bibnamefont{Slijper}},
  \bibinfo{author}{\bibfnamefont{R.}~\bibnamefont{Boelens}},
  \bibinfo{author}{\bibfnamefont{A.~L.} \bibnamefont{Davis}},
  \bibinfo{author}{\bibfnamefont{R.~N.~H.} \bibnamefont{Konings}},
  \bibinfo{author}{\bibfnamefont{G.~A.} \bibnamefont{van~der Marel}},
  \bibinfo{author}{\bibfnamefont{J.~H.} \bibnamefont{van Boom}},
  \bibnamefont{and} \bibinfo{author}{\bibfnamefont{R.}~\bibnamefont{Kaptein}},
  \bibinfo{journal}{Biochemistry} \textbf{\bibinfo{volume}{36}},
  \bibinfo{pages}{249} (\bibinfo{year}{1997}), \bibinfo{note}{-PDB ID: 1LQC}.

\bibitem[{\citenamefont{Barkley and Bourgeois}(1980)}]{BarkBour80}
\bibinfo{author}{\bibfnamefont{M.~D.} \bibnamefont{Barkley}} \bibnamefont{and}
  \bibinfo{author}{\bibfnamefont{S.}~\bibnamefont{Bourgeois}},
  \emph{\bibinfo{title}{The Operon}} (\bibinfo{publisher}{Cold Spring Harbour},
  \bibinfo{year}{1980}), chap.~\bibinfo{chapter}{7}.

\bibitem[{\citenamefont{Horton et~al.}(1997)\citenamefont{Horton, Lewis, and
  Lu}}]{HortLewLu97}
\bibinfo{author}{\bibfnamefont{N.}~\bibnamefont{Horton}},
  \bibinfo{author}{\bibfnamefont{M.}~\bibnamefont{Lewis}}, \bibnamefont{and}
  \bibinfo{author}{\bibfnamefont{P.}~\bibnamefont{Lu}},
  \bibinfo{journal}{Journal of molecular biology}
  \textbf{\bibinfo{volume}{265}}, \bibinfo{pages}{1} (\bibinfo{year}{1997}).

\bibitem[{\citenamefont{Levandoski et~al.}(1996)\citenamefont{Levandoski,
  Tsodikov, Frank, Melcher, Saecker, and {Record Jr}}}]{Leva+96}
\bibinfo{author}{\bibfnamefont{M.~M.} \bibnamefont{Levandoski}},
  \bibinfo{author}{\bibfnamefont{O.~V.} \bibnamefont{Tsodikov}},
  \bibinfo{author}{\bibfnamefont{D.~E.} \bibnamefont{Frank}},
  \bibinfo{author}{\bibfnamefont{S.~E.} \bibnamefont{Melcher}},
  \bibinfo{author}{\bibfnamefont{R.~M.} \bibnamefont{Saecker}},
  \bibnamefont{and} \bibinfo{author}{\bibfnamefont{M.~T.} \bibnamefont{{Record
  Jr}}}, \bibinfo{journal}{Journal of Molecular Biology}
  \textbf{\bibinfo{volume}{260}}, \bibinfo{pages}{697} (\bibinfo{year}{1996}).

\bibitem[{\citenamefont{Brunger et~al.}(1998)\citenamefont{Brunger, Adams,
  Clore, Delano, Gros, Grosse-Kunstleve, Jiang, Kuszewski, Nilges, Pannu
  et~al.}}]{cns}
\bibinfo{author}{\bibfnamefont{A.}~\bibnamefont{Brunger}},
  \bibinfo{author}{\bibfnamefont{P.}~\bibnamefont{Adams}},
  \bibinfo{author}{\bibfnamefont{G.}~\bibnamefont{Clore}},
  \bibinfo{author}{\bibfnamefont{W.}~\bibnamefont{Delano}},
  \bibinfo{author}{\bibfnamefont{P.}~\bibnamefont{Gros}},
  \bibinfo{author}{\bibfnamefont{R.}~\bibnamefont{Grosse-Kunstleve}},
  \bibinfo{author}{\bibfnamefont{J.-S.} \bibnamefont{Jiang}},
  \bibinfo{author}{\bibfnamefont{J.}~\bibnamefont{Kuszewski}},
  \bibinfo{author}{\bibfnamefont{N.}~\bibnamefont{Nilges}},
  \bibinfo{author}{\bibfnamefont{N.}~\bibnamefont{Pannu}},
  \bibnamefont{et~al.}, \bibinfo{journal}{ACTA CRYST.}
  \textbf{\bibinfo{volume}{D54}}, \bibinfo{pages}{905} (\bibinfo{year}{1998}).

\bibitem[{\citenamefont{Yildirim and Mackey}(2003)}]{Yild+03}
\bibinfo{author}{\bibfnamefont{N.}~\bibnamefont{Yildirim}} \bibnamefont{and}
  \bibinfo{author}{\bibfnamefont{M.~C.} \bibnamefont{Mackey}},
  \bibinfo{journal}{Biophys. J.} \textbf{\bibinfo{volume}{84}},
  \bibinfo{pages}{2841} (\bibinfo{year}{2003}).

\bibitem[{\citenamefont{Spiegel et~al.}(1991)\citenamefont{Spiegel, {Backlund
  Jr}, Butrynski, Jones, and Simonds}}]{Spie+91}
\bibinfo{author}{\bibfnamefont{A.~M.} \bibnamefont{Spiegel}},
  \bibinfo{author}{\bibfnamefont{P.~S.} \bibnamefont{{Backlund Jr}}},
  \bibinfo{author}{\bibfnamefont{J.~E.} \bibnamefont{Butrynski}},
  \bibinfo{author}{\bibfnamefont{T.~L.} \bibnamefont{Jones}}, \bibnamefont{and}
  \bibinfo{author}{\bibfnamefont{W.~F.} \bibnamefont{Simonds}},
  \bibinfo{journal}{TIBS} \textbf{\bibinfo{volume}{16}} (\bibinfo{year}{1991}).

\bibitem[{\citenamefont{Swint-Kruse et~al.}(2003)\citenamefont{Swint-Kruse,
  Zhan, Fairbanks, Maheshwari, and Matthews}}]{Swin+03}
\bibinfo{author}{\bibfnamefont{L.}~\bibnamefont{Swint-Kruse}},
  \bibinfo{author}{\bibfnamefont{H.}~\bibnamefont{Zhan}},
  \bibinfo{author}{\bibfnamefont{B.~M.} \bibnamefont{Fairbanks}},
  \bibinfo{author}{\bibfnamefont{A.}~\bibnamefont{Maheshwari}},
  \bibnamefont{and} \bibinfo{author}{\bibfnamefont{K.~S.}
  \bibnamefont{Matthews}}, \bibinfo{journal}{Biochemistry}
  \textbf{\bibinfo{volume}{42}}, \bibinfo{pages}{14004} (\bibinfo{year}{2003}).

\end{thebibliography}
\end{document}